\documentclass[twocolumn,trackchanges]{aastex701}
\usepackage{caption}

\usepackage{amsmath}
\usepackage{cleveref}
\usepackage{subcaption}
\usepackage{float}
\usepackage{xcolor}

\begin{document}
	
	\title{\large  Relativistic Dynamical Evolution of Kilonova Spectral Emission}
	
	\author[orcid=0000-0000-0000-0001,sname='Karimi']{Sara Karimi}
	\affiliation{Department of Physics, Shahid Beheshti University, Tehran 19839-63113, Iran}
	\email[show]{\url{sarakarimi2002@gmail.com}}
	
	\author[orcid=0000-0002-1765-0494,sname='Shakeri']{Soroush Shakeri}
	\affiliation{Department of Physics, Isfahan University of Technology, Isfahan 84156-83111, Iran}
	\email[show]{\\ \url{s.shakeri@iut.ac.ir}}

	\begin{abstract}
		
	We develop a relativistic dynamical model for the thermal evolution of expanding kilonova ejecta, in which the effective temperature is determined by the ejecta mass, opacity, velocity, and radioactive heating rate. The resulting thermal evolution follows a broken power law as the ejecta transition from the optically thick to the optically thin regime. We incorporate this thermal history into a relativistic spectral calculation that accounts for the equal-arrival-time surface geometry and Doppler boosting, enabling us to model the observed kilonova spectrum and constrain the underlying ejecta properties. We investigate the effects of the model parameters and relativistic corrections on the observed temperature and spectral shape, and find that light-travel-time delays and Doppler shifts can produce significant spectral distortions at high ejecta velocities. We apply the model to the early-time X-shooter spectrum of the kilonova AT2017gfo and infer the ejecta mass, opacity, and velocity through Markov chain Monte Carlo (MCMC) fitting with a Student's $t$ likelihood. We find that relativistic effects become increasingly important for fast ejecta ($v \gtrsim 0.5c$), particularly at large viewing angles, where the observed temperature can differ substantially from the rest-frame effective temperature. These results highlight the importance of incorporating relativistic and equal-arrival-time effects when interpreting early-time kilonova spectra, especially for rapidly expanding ejecta.
		
	\end{abstract}

	\section{ Introduction}
	
	{\bf \it \large  Kilonovae} are radioactively powered electromagnetic transients produced a the neutron-rich ejecta of binary neutron star  mergers. The discovery of AT2017gfo in association with GW170817 provided the first direct observational confirmation that such events are major sites of rapid neutron-capture ($r$-process) nucleosynthesis \citep{Smartt_2017,2017PhRvL}. The early-time spectra of AT2017gfo, despite the expected complexity of the expanding merger ejecta, were found to be remarkably well described by 
	the blackbody continuum originating from thermalization caused by radioactive decay of heavy elements, \citep{2017Natur.551...67P, 2017Sci...358.1570D, 2017Sci...358.1574S, 2017ApJ...848L..32M,Waxman:2019png,10.1093/mnras/sty2441, 2019Natur.574..497W}. 
	
	There have been numerous theoretical attempts to model the kilonova emissions and evaluate the   physical properties of the ejecta material and its composition \citep{arnet,Cowperthwaite2017,Kasen_2017,Villar2017,Metzger:2010sy,Metzger_2017}. These models rely on simplifying assumptions, such as the diffusion limit, homogeneous energy deposition, and a uniformly expanding medium represented by one or more effective components (e.g.,\ a low-opacity blue component and a higher-opacity red component). The kilonova radiation is also approximated as  blackbody emission, and radioactive heating contributes to the thermalization of the ejecta \citep{Khatami:2018dbu, Kasen_2019,Sukhbold:2018hsg,Ricigliano:2023svx}.  
	
	For more accurate modeling of the early-time emission of the kilonova when  the ejecta is still optically thick, one must solve the radiative transfer equations accounting for wavelength- and time-dependent opacities and viewing-angle-dependent ejecta geometries, which is possible with numerical radiation-hydrodynamics simulations \citep{Wollaeger2018,Bulla2019,Korobkin2021,Chen:2021tob}.
	
	Thus, kilonova modeling has progressed from relatively simple diffusion-based analytical and semi-analytical prescriptions to increasingly complex numerical radiation-transport calculations that couple ejecta dynamics, nuclear heating, opacity, geometry, and viewing-angle effects. 
	Detailed numerical radiative transfer simulations, while accurate, require high computational cost and are often too complex to provide simple physical intuition about kilonovae. Intermediate approaches have also been developed that provide
	effective descriptions linking observations to theoretical models and numerical
	simulations \citep{Liu:2025grx,Kitamura:2025auo}.

	A semi-analytic approach was developed in \citep{Li_1998,Li_2019}, in which the dynamical and thermal evolution of the kilonova ejecta was obtained by solving the internal-energy equation, including radioactive heating, adiabatic cooling, and radiative diffusion. This framework was primarily used to model the bolometric light curve and to determine the evolution of the  effective temperature of the ejecta in the frame of the explosion center. 
	However, the thermal radiation received by a distant observer, particularly from rapidly expanding ejecta, is additionally affected by geometrical light-travel-time effects and relativistic corrections such as Doppler boosting. While these effects can modify the observed radiation, they are largely integrated out in the case of the bolometric emission.

	The effects of the equal-arrival-time surface (EATS; \citealt{Zhang_2018}) and Doppler shifting on the variation of apparent temperature were addressed by \citep{Sneppen_2023,Sneppen:2023vkk}, where it was shown that the observed kilonova spectrum of relativistic ejecta is,
	in general, a superposition of blackbody spectra at different apparent temperatures, giving rise to an effectively multi-temperature blackbody spectrum. It was also shown that  
	neglecting relativistic effects can introduce systematic errors of up to $\sim50\%$ in the inferred temperature and velocity
	in the case of mildly relativistic ejecta \cite{Sadeh:2025oez}. Although these studies incorporated the relevant geometrical and relativistic effects, the emitted temperature was prescribed phenomenologically as a power law rather than being derived self-consistently from the ejecta dynamical evolution.

	In this work, we present a dynamical model to describe the thermal evolution of the ejecta in terms of the ejecta mass, opacity, velocity, and the radioactive heating rate. The resulting temperature exhibits a broken power-law behavior as the ejecta transitions from the optically thick to the optically thin regime. We then incorporate this thermal history into a relativistic spectral calculation based on the EATS geometry and Doppler boosting to compute  the observed spectrum, enabling the physical properties of the ejecta to be inferred from kilonova observations. We find that, at relativistic ejecta velocities, light-travel-time delays and Doppler shifting significantly modify the observed emission, causing spectral distortions. We then consider the impact of various model parameters on the kilonova spectra and finally compare  the model predictions with the observed spectrum of the event AT2017gfo.

	The present paper is organized as follows. In Section~\ref{sec:thermalevol}, we establish the dynamical model to evaluate the thermal evolution of the expanding ejecta in its rest frame. In Section~\ref{sec:relativistic}, we introduce the relativistic and geometrical corrections into the model and consider their effects on the observed temperatures and the spectrum. In Section~\ref{sec:synthetic}, we focus on the kilonova event AT2017gfo taking into account the early-time X-shooter spectra in order to infer the ejecta parameters via Markov Chain Monte Carlo (MCMC) fitting by adopting Student-t likelihood. We finally summarize our main conclusions and give some concluding remarks in Section~\ref{sec:conclusion}.

	\section{Thermal Energy Evolution of the Expanding
		Ejecta} \label{sec:thermalevol}

	According to the first law of thermodynamics, the internal-energy evolution of the expanding ejecta is driven by radioactive heating from decaying $r$-process elements, adiabatic expansion, and radiative losses, and can be written as follows \citep{1980ApJ...237..541A,arnet,2017ApJ...846...33A}:
	\begin{equation}
		\frac{dE_{\mathrm{int}}}{dt} = \dot{Q}_{\mathrm{in}} - L - P \frac{dV}{dt}
		\label{eq:main}
	\end{equation}
	where $E_{\mathrm{int}} $ is the total internal energy, $\dot{Q}_{\mathrm{in}}$ is the heating rate due to radioactive decay, and $L$ is the radiative power. The last term represents the work performed by adiabatic expansion \citep{Hotokezaka_2020}.
	
	Here, we consider the merger ejecta as a spherically symmetric, homogeneous sphere of total mass $M_{\rm ej}$ expanding at a constant velocity $v$ at its surface \citep{Li_1998,Li_2019}. The radius of the expanding sphere is then $R_{\rm ej}(t) = vt$, so the mass density decreases as
	, so the mass density decreases as
	\begin{equation}
		\rho(t)=\frac{3M_{\rm ej}}{4\pi R_{ej}^3}\approx\frac{3M_{\rm ej}}{4\pi v^3}\,t^{-3}.
		\label{eq:rho_homol}
	\end{equation}
	In the early phase, the ejecta is hot and optically thick ($\tau_s\gg1$), and the internal energy density $U$  is dominated by the radiation, so that $U = 3P= aT^4$ where $T$ is the radiation temperature and $a$ is the radiation constant. But rapid expansion eventually causes the ejecta to become optically thin, releasing the thermal emission from the radioactive decay.
	Assuming a grey (frequency–independent) opacity $\kappa$ that is spatially uniform, the optical depth at the photospheric radius is
	
	\begin{equation}\label{e3}
		\tau(R_{ph})=\int_{R_{ph}}^{R_{ej}} \rho \kappa dr = \rho \kappa (R_{ej}-R_{ph}).
	\end{equation}
	Assuming negligible material beyond the ejecta, we define the photospheric radius by $\tau(R_{ph}) \approx 1$. Rearranging Eq. (\ref{e3}), $R_{ph}$ is given by  
	
	\begin{equation}
		R_{ph} = R_{ej}-\frac{1}{\kappa \rho(t)},
		\label{eq:photosphere}
	\end{equation}
	
	At early times, when $\kappa\rho$ is large, $R_{\rm ph} \approx R_{\rm ej}$, and the photosphere lies very close to the surface of the ejecta. Independent of the density distribution, $R_{\rm ph}$ increases at early times and decreases at late times \cite{Liu:2018hft}.  As the ejecta expands, its total optical depth decreases as
	$\kappa \rho(t) R_{\rm ej} \propto t^{-2}$, and after a characteristic
	timescale $t_c$, the ejecta becomes optically thin
	\begin{equation}
		t_c=\left(\frac{3\kappa M_{\rm ej}}{4\pi v^{2}}\right)^{1/2}.
		\label{critical}
	\end{equation}
	
	The expression for $R_{\rm ph}$ (Eq. \ref{eq:photosphere}) is physically meaningful only while $\kappa \rho(t) R_{\rm ej} \gtrsim 1$. For $t \gtrsim t_c$, the total
	optical depth falls below unity and a well-defined photospheric surface
	no longer exists. As in the one-zone approach of \citep{Li_1998,Li_2019}, we approximate the effective emitting surface as $R_{\rm ej}(t)$, and the transition from the optically thick to the optically thin regime is incorporated through the effective temperature $T_{\rm eff}$ :
	
	\begin{equation}\label{e6}
		T_{\mathrm{eff}}^4 \simeq \frac{T^4}{\tau_s + 1/2},
	\end{equation}
	which depends explicitly on the total optical depth.
	
	In terms of $t_c$, the total optical depth  can be written  as
	\begin{equation}
		\tau_s(t)=\left(\frac{t}{t_c}\right)^{-2}.
		\label{eq:tau_tc}
	\end{equation}
	
	\begin{figure*}[!htbp]
		\centering
		\includegraphics[width=0.8\linewidth]{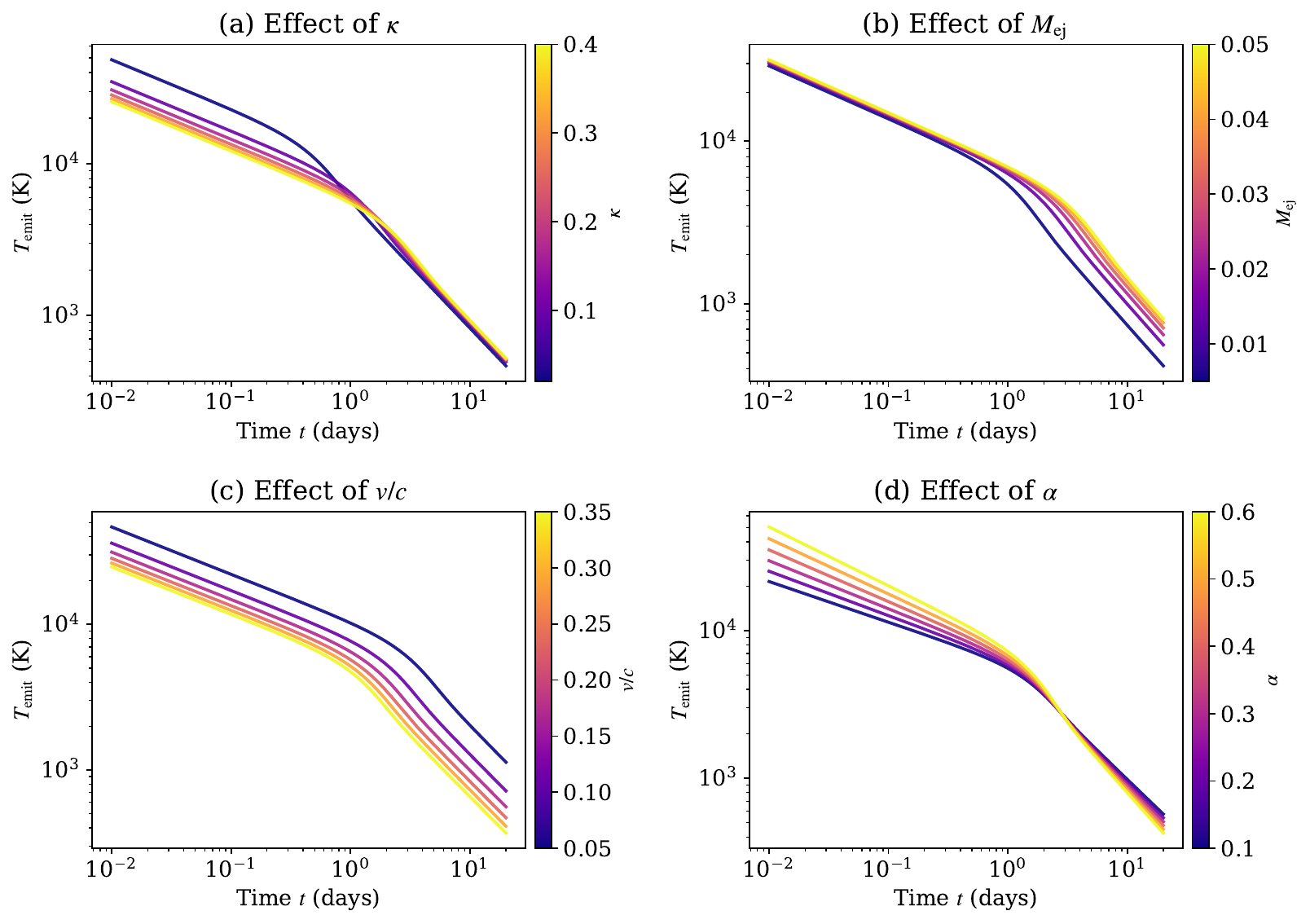}
		\caption{The time evolution of the emitted temperature ($T_{\rm emit}$) in the dynamical model presented in Section \ref{sec:thermalevol} for different values of the ejecta parameters, including the mass $M_{\rm ej}$, velocity $\beta=v/c$, opacity $\kappa$, and radioactive heating index $\alpha$, as labeled.}
		\label{fig:li_dynamic}
	\end{figure*}

	Returning to Eq. \eqref{eq:main}, the heating rate of radioactive decay is given by $\dot{Q}_{\mathrm{in}} = M_{\rm ej} \epsilon(t)$ \cite{Li_1998} where $\epsilon(t)$ is the specific heating rate and as in \cite{Li_2019},
	we adopt a general power-law form 
	\begin{equation}
		\epsilon(t) = \frac{f c^2}{t_c} 
		\left(\frac{t_c}{t}\right)^{1+\alpha},
		\label{eq:heating_rate}
	\end{equation}
	where $f$ and $\alpha$ are dimensionless parameters which are related to the average energy released by radioactive decay of r-process elements within the ejecta. The brightness of the kilonova event depends crucially on the amount of radioactive heating energy  which contributes to the peak luminosity \citep{Metzger:2010sy}. In fact, it was shown that the radioactive decay from a large statistical sample of isotopes can generate a heating rate which can be well described by a single power-law as in Eq. (\ref{eq:heating_rate})  \cite{Li_1998,Metzger:2010sy}.

	
	
	The efficiency parameter $f$ can be derived from the underlying distribution of radioactive nuclides. Rather than a single heating-efficiency parameter, more accurate kilonova modeling requires the incorporation of a comprehensive nuclear reaction network to properly account for the relevant nuclear processes \citep{10.3389/fspas.2023.1243615,pynucastro_development_team_2025_17455462}.
	
	\begin{figure*}[t]
		\centering
		\includegraphics[width=0.75\linewidth]{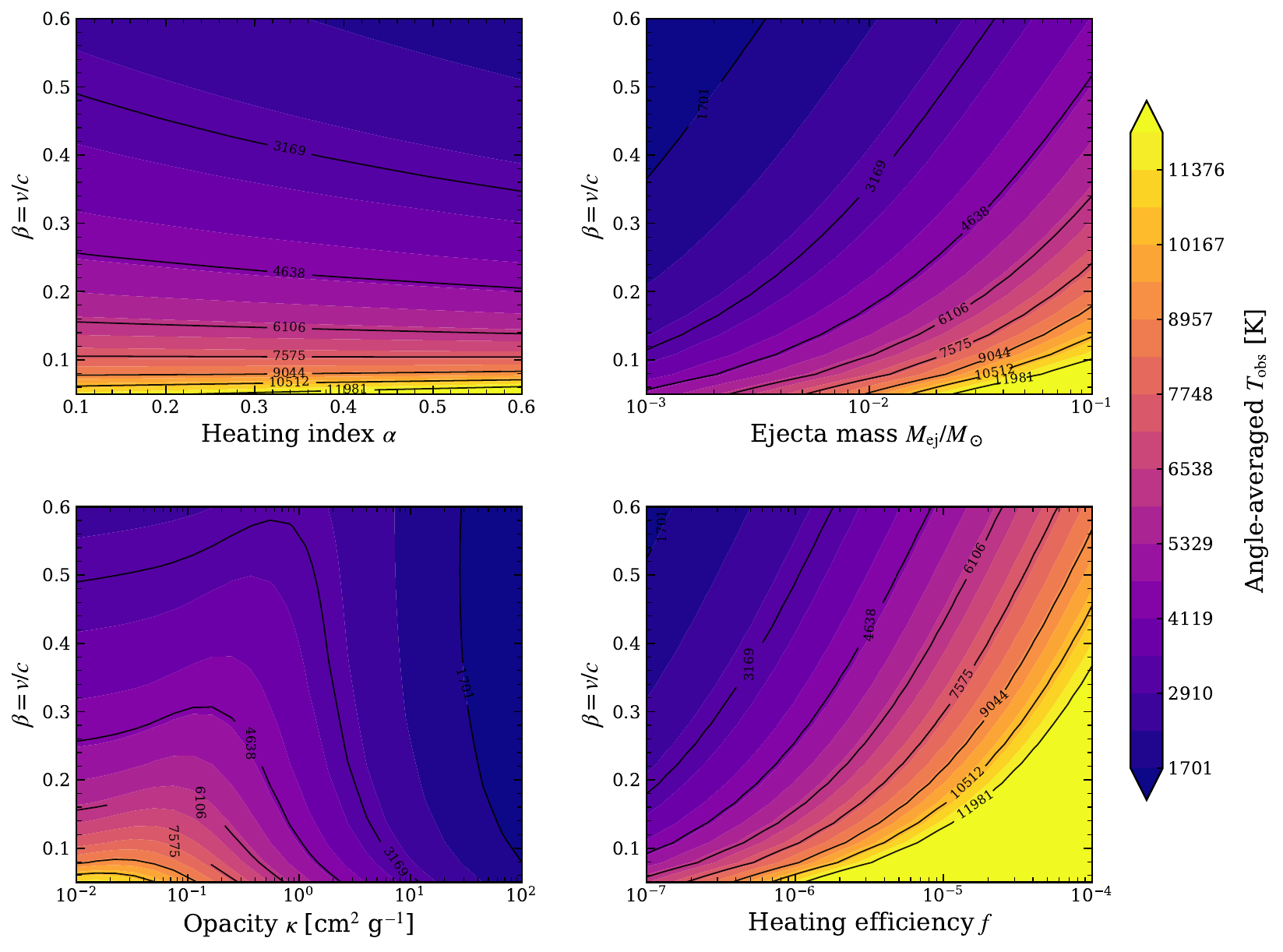}
		\caption{Dependence of the angle-averaged observed temperature at 1.43 days after merger on the  ejecta parameters.
			Each panel shows $T_{\text{eff,obs}}$ as a function of the velocity $\beta=v/c$ and one additional parameter: (top left) heating index $\alpha$, (top right) ejecta mass $M_{\rm ej}$, (bottom left) opacity $\kappa$, and (bottom right) heating efficiency $f$.}
		\label{fig:param_dependence}
	\end{figure*}
	
	A fraction of $\dot{Q}_{\mathrm{in}}$ is trapped by the ejecta and consequently thermalized, while the remainder escapes as $\gamma$-ray photons \citep{Li_2019,2016ApJ...829..110B}. The thermalized fraction is approximated by $1 - e^{-\tau_s}$, with $\tau_s$ being the optical depth to $\gamma$-rays \citep{Kasen_2019}.
	Following \citep{Li_1998,Li_2019} for the simplicity,  we adopt the the same grey opacity to approximate both the optical and $\gamma$-ray thermalization. Thus, the deposited heating can be expressed as
	\begin{equation}
		\dot{Q}_{\mathrm{in}}^{\mathrm{th}} = (1 - e^{-\tau_s}) \dot{Q}_{\mathrm{in}} 
		= \left( 1 - e^{-t_c^2 / t^2} \right)\epsilon(t) M_{\rm ej},
		\label{eq:heating}
	\end{equation}
	In order to evaluate Eq.~(\ref{eq:main}), we take the bolometric luminosity $L$ to be

	\begin{equation}
		L = 4\pi R^2 F = \left( \frac{4\pi^2 v^4 c}{3 \kappa M_{\rm ej}} \right) \frac{U t^4}{1 + t^2 / (2 t_c^2)}
		\label{eq:luminosity}
	\end{equation}
	
	We identify the emitting radius as $R \equiv R_{\rm ej}(t)=vt$ and the flux is approximated as $F = \sigma T_{\mathrm{eff}}^4$ in which the effective temperature $T_{\mathrm{eff}}$ is given by Eq. (\ref{e6}) to ensure a smooth bridging from optically thick to the optically thin 
	regime \citep{Carroll_Ostlie_2017,Li_2019}.
	Using Eqs.~(\ref{eq:heating}), and (\ref{eq:luminosity}), Eq.~(\ref{eq:main}) becomes

	\begin{equation}
		\begin{split}
			\frac{d}{dt} \left( t^4 U \right)
			&= \left( \frac{3M_{\rm ej}}{4\pi v^3} \right) \epsilon(t)\, t\, 
			\big(1 - e^{-t_c^2/t^2}\big)
			\\
			&\quad
			- \left( \frac{\pi v c}{\kappa M_{\rm ej}} \right) 
			\frac{t^5 U}{1 + t^2 / (2t_c^2)}
		\end{split}
		\label{eq:final}
	\end{equation}
	
	where $U$ is the energy density of the radiation. The solution of Eq. (\ref{eq:final}) describes the dynamical evolution of the energy density of the kilonova, from which the bolometric luminosity follows via Eq. (\ref{eq:luminosity}). 
	While the bolometric light curve captures the overall brightness, the spectral energy distribution (SED) encodes the information about the ejecta’s composition. 
	In the following, we aim to include the relativistic corrections in the dynamical model to reproduce the observed  kilonova spectrum.

	\section{Relativistic Effects \& Geometrical Corrections} \label{sec:relativistic}

	Here, we incorporate relativistic and geometric effects by mapping this dynamical temperature solution onto the equal-arrival-time surface (EATS; \cite{Zhang_2018}). For a surface element at angle \(\theta\) (where \(\mu = \cos\theta\)), the emission time ($t$) in the rest-frame is related to a single observer time \(t_{\mathrm{obs}}\) as
	\begin{equation}
		t(\mu, t_{\mathrm{obs}}) = \frac{t_{\mathrm{obs}}}{(1 - \beta\mu)}.
		\label{eq:delay}
	\end{equation}
	This relation describes the geometric light-travel–time delay, allowing photons emitted at different angles and times to arrive at the observer simultaneously.
	
	Substituting 
	Eq.~(\ref{eq:delay}) into the homologous expansion law $R(t)=\beta c\,t$ yields 
	the equal--arrival--time radius, $R(\mu, t_{\rm obs}) = \tfrac{\beta c t_{\rm obs}}{1 - \beta \mu}$. Here, we assume emission from the surface of a single expanding sphere. We normalize $R(\mu, t_{\rm obs})$ by the emitting line–of–sight radius $R(\mu =1, t_{\rm obs})$ as
	\begin{equation}
		\frac{R(\mu, t_{\rm obs})}{R(\mu=1, t_{\rm obs})} = 
		\frac{1-\beta}{1-\beta\mu},
	\end{equation}
	In our formalism, we adopt the radiative diffusion limit, and the ejecta's optical-depth evolution is incorporated into the effective temperature through Eq. \ref{e6}.

	In the dynamical model presented in Sec.~\ref{sec:thermalevol}, the time evolution of the emitting temperature \( T_{\mathrm{ emit}}(t) \) ($ T_{\mathrm{emit}}(t) \equiv T_{\mathrm{eff}}(t)$) is determined by solving  the internal energy density  Eq.~(\ref{eq:final})  in the ejecta's rest-frame and using  Eq.~(\ref{eq:luminosity}). The thermal evolution depends  
	on the physical  parameters of the kilonova  such as the ejecta mass, velocity, opacity  and the heating index $\alpha$ as presented in  Fig.~\ref{fig:li_dynamic}. Each panel shows the time evolution of the emitted temperature $T_{\rm emit}$ in the dynamical model  over $0.01\leq t\leq20$ days, where one parameter is varied (indicated by the color bar), while the rest are fixed at $M_{\rm ej}=0.01\,M_\odot$, $\kappa=0.2~{\rm cm^2\,g^{-1}}$, $\alpha=0.3$, $\beta=0.2$ and $f=10^{-6}$.

	We see that  \( T_{\mathrm{emit}}(t) \) shows a broken power-law behaviour, where the break occurs at the diffusion timescale \( t_c \) (depending on  $M_{ej}$, $v$ and $\kappa$ through Eq.  (\ref{critical})), when the ejecta becomes transparent. Note that we can also obtain analytical solutions of Eq. (\ref{eq:final}) in the limiting cases of optically thick and optically thin ejecta, which yield the temporal evolution of the effective temperature as
	\begin{equation}
		T_{\mathrm{emit}}(t) \propto
		\begin{cases}t^{-(1+\alpha)/4} \ \ \ \ t\ll t_{c}
			\\ t^{-(5+\alpha)/4} \ \ \ \ t\gg t_{c}
		\end{cases}\textcolor{green}{.}\label{e15}
	\end{equation}
	As can be seen, $T_{\mathrm{emit}}$ decays  faster in the optically thin regime ($t\gg t_{c}$) and the heating index $\alpha$ which appears in the power law exponents originates from the time-dependence of the radioactive energy input 
	($\dot{Q}\propto t^{-(1+\alpha)}$) (See Eq. (\ref{eq:heating_rate})).

	Moreover, the observed temperature of each surface element is 
	Doppler-shifted,
	\begin{equation}
		T_{\rm obs}(\mu,t_{\rm obs}) = \frac{T_{\rm emit}(t(\mu,t_{\rm obs}))
		}{\gamma(1-\beta\mu)}.
		\label{e14}
	\end{equation}
	This shows that taking into account both the light-travel-time delay effect and Doppler shift leads to an angular dependence of $T_{\text{obs}}$,  reflecting  a multi-temperature characteristic of the relativistic expanding ejecta. To characterize the global thermal behavior, 
	we compute the angular average of the observed temperature  $T_{\text{eff,obs}}$, which provides a single 
	effective temperature to describe the radiative surface. In order to define a single effective blackbody temperature $T_{\text{eff,obs}}$, we use  the bolometric luminosity as
	
	\begin{equation}\label{eee16}
		L_{\rm bol}(t_{\rm obs}) = \int_{0}^{\infty} L_\lambda(t_{\rm obs})\, d\lambda
		= A\,\sigma\,T_{\rm eff,obs}^{4},
	\end{equation} 
	where the corresponding effective emitting area becomes
	
	\begin{equation}\label{eee17}
		A = 8\pi \int_{\beta}^{1} R^{2}(\mu,t_{\text{obs}}) \mu d \mu,
	\end{equation} 
	and
	\begin{multline}\label{ee17}
		L_\lambda(t_{\rm obs}) = 
		8\pi^2\int_{\beta}^{1} 
		B_\lambda\!\left[T_{\rm obs}(\mu,t_{\rm obs})\right]  R^{2}(\mu,t_{\rm obs})
		\mu\, d\mu .
	\end{multline}

	\begin{figure*}[!htbp]
		\vspace{-2mm} 
		\centering
		\includegraphics[width=1\linewidth]{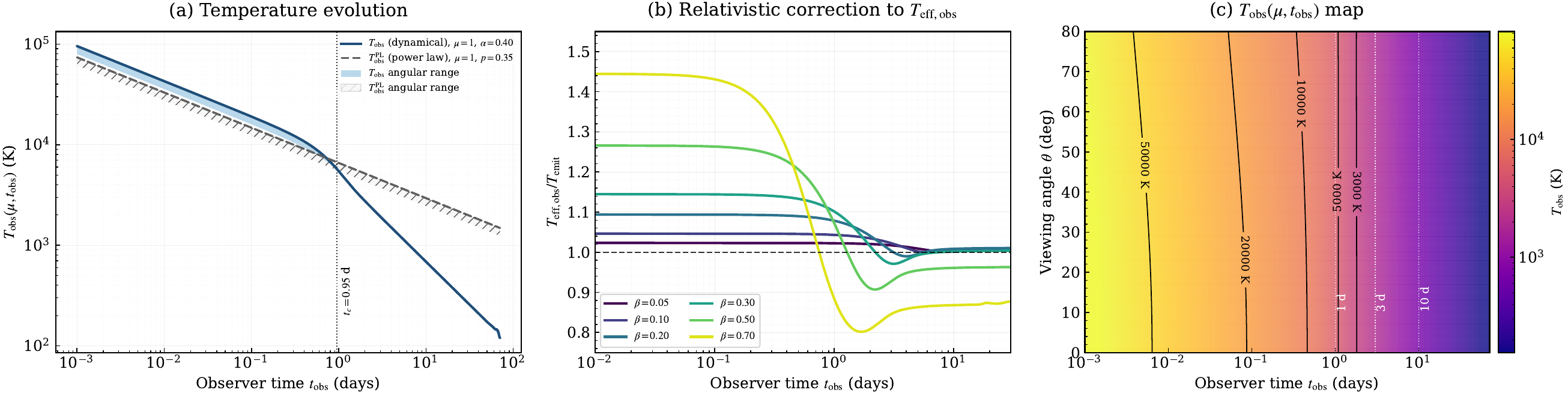}
		\vspace{-1mm} 
		\caption{
			The evolution of the observer-frame temperatures for the ejecta with 
			$M_{\rm ej}=0.01\,M_\odot$ and $\kappa=0.1~{\rm cm^2\,g^{-1}}$. Panel~(a) shows the baseline evolution for the observed temperature in line-of-sight viewing and the angular spread for the dynamical model as well as a simple power law  $T_{\rm emit}^{\rm PL}(\mu,t_{\rm obs})\propto[t_{\mathrm{obs}}(1-\beta\mu)]^{-p}$ for $\beta=0.28$,  panel~(b)  the Doppler effect on the  ratio of the angular average of the temperature to emitted temperature $T_{\rm eff, obs}/T_{\rm emit}$ for various values of $\beta$ as labeled,
			and  panel~(c) presents the heat map $T_{\mathrm{obs}}(t_{\mathrm{obs}},\mu)$  and selected contours indicate some specific temperatures  }
		
		\label{fig:Tobs_comparison}
		\vspace{2mm} 
	\end{figure*}

	In the above expression, we integrate the delayed and Doppler–shifted blackbody contributions over the emitting surface to obtain the observer specific luminosity  $L_\lambda(t_{\rm obs})$.
	In fact, the observed spectrum is not a 
	single blackbody but a superposition of Doppler-shifted blackbodies 
	originating from different latitudes on the emitting surface and different emission times. Observed temperature has angular dependence through Eq.~(\ref{e14}), and therefore, the observed spectrum is intrinsically multi–temperature.

	In Fig.~\ref{fig:param_dependence},  we display the results of a scan over the parameter space of kilonova ejecta  in order to illustrate the impact of the relativistic corrections on $ T_{\text{eff}, \rm obs}$ at $t_{\rm obs}= 1.43$ days after the merger. 
	In each panel, the vertical axis shows  the variation of the ejecta velocity  $\beta = v/c$ and the horizontal axis is one of the ejecta parameters, while the other parameters are fixed at values $M_{\rm ej}=0.021\,M_\odot$, $\kappa=0.01~{\rm cm^2\,g^{-1}}$, $\alpha=0.1$ and $f=10^{-6}$. 
	
	\begin{figure*}[t]
		\centering
		\includegraphics[width=\textwidth]{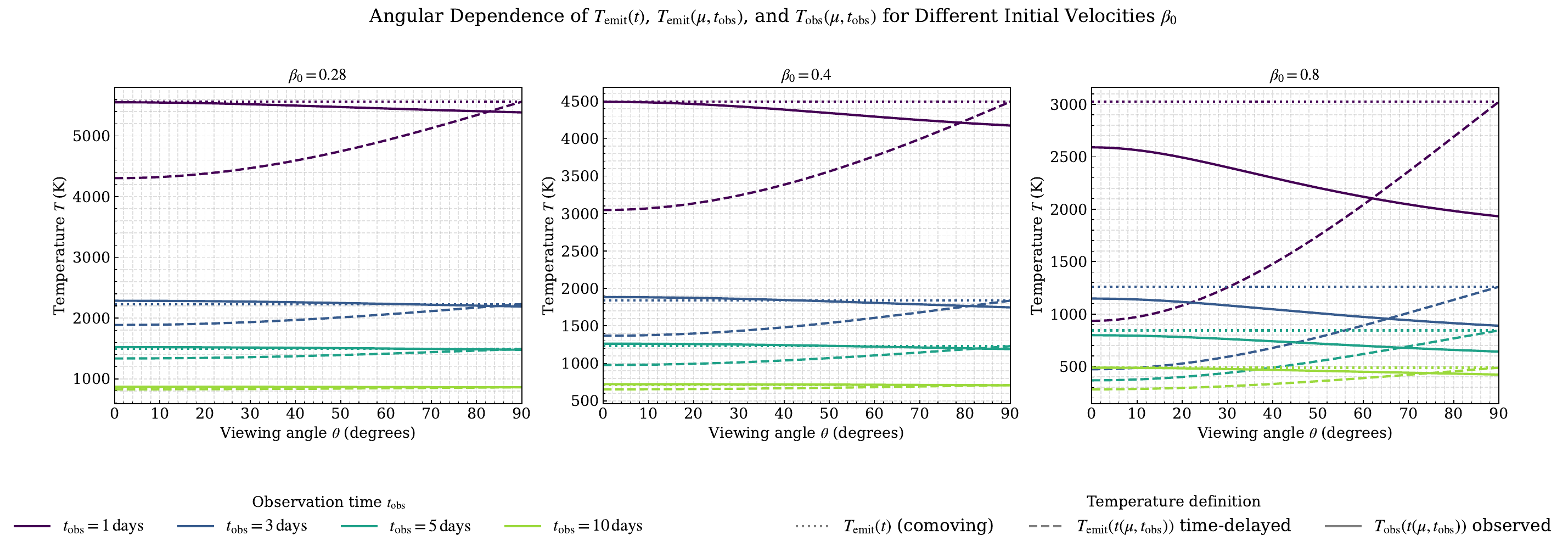}
		\caption{Angular dependence of the emitting, time-delayed, and observed temperatures 
			for different ejecta velocities $\beta_0$. 
			Each panel corresponds to a distinct expansion velocity 
			($\beta_0=0.28$, $0.4$, and $0.8$), showing 
			$T_{\mathrm{emit}}(t)$ (dotted), 
			$T_{\mathrm{emit}}(t(\mu,t_{\mathrm{obs}}))$ (dashed), and 
			$T_{\mathrm{obs}}(t(\mu,t_{\mathrm{obs}}))$ (solid) 
			at several observation times 
			($t_{\mathrm{obs}}=1$, 3, 5, and 10 d). 
			The model parameters used in this figure are 
			$\kappa=0.022 ~{\rm cm^2\,g^{-1}}$, 
			$M_{\rm ej}=0.016\,M_\odot$, 
			$\alpha=0.166$, and 
			$f=10^{-6}$.
			For mildly relativistic ejecta ($\beta_0\lesssim0.3$), 
			the time-delay and Doppler effects act in opposite directions: 
			the delay increases the limb temperature, while Doppler boosting 
			enhances the temperature near $\mu\approx1$. 
			Their near-cancellation results in a nearly uniform 
			$T_{\mathrm{obs}}(t_{\mathrm{obs}},\mu)$, explaining why AT2017gfo appears 
			approximately as a single-temperature blackbody despite its 
			expanding geometry.
		}
		\label{fig:Teff_angular_dependence}
	\end{figure*}
	
	\begin{figure*}[!htbp]
		\centering
		\includegraphics[width=.8\linewidth]{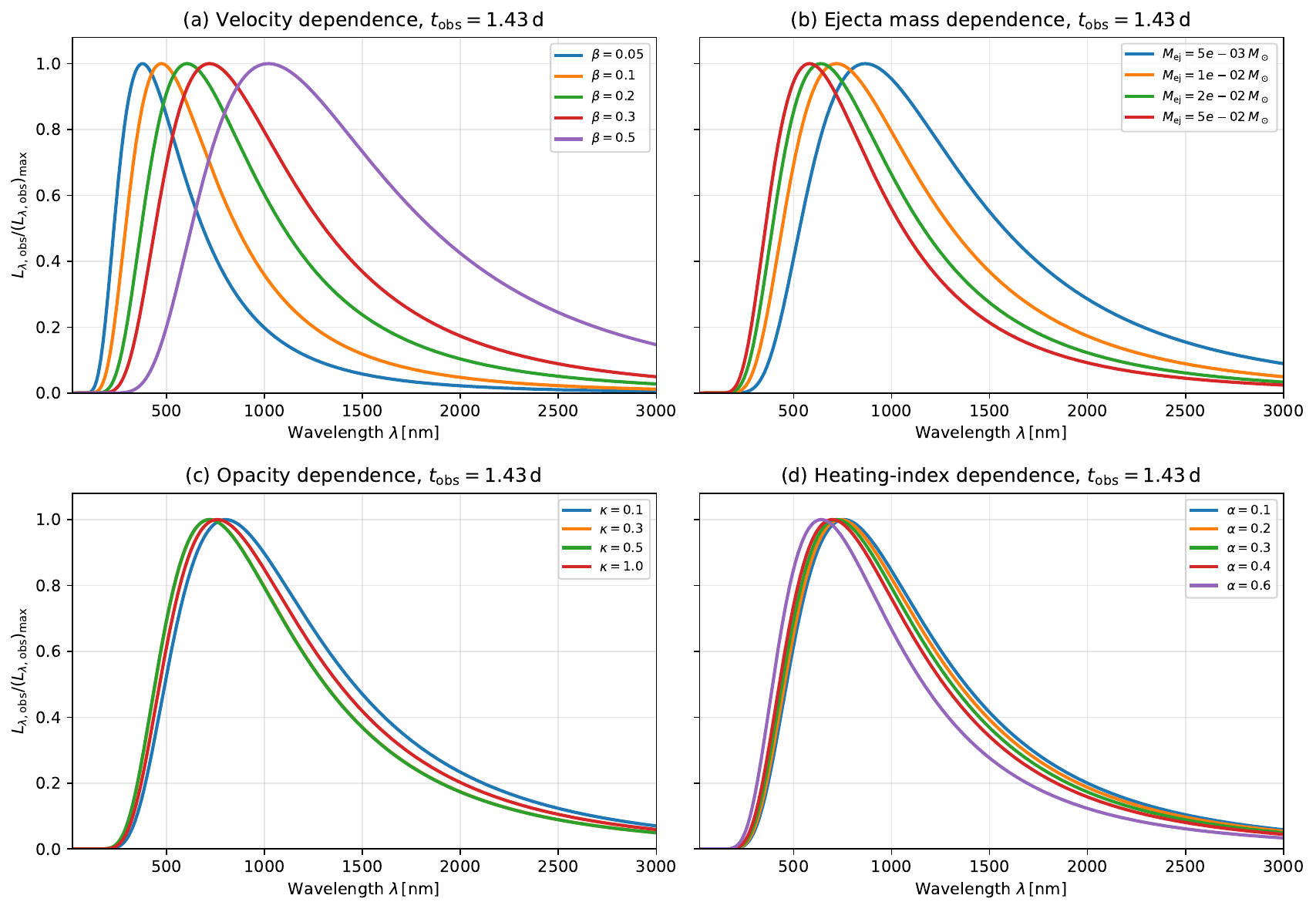}
		
		\caption{The normalized kilonova spectra for 1.43 days after merger considering the impact of the  various parameters such as the ejecta mass $M_{ej}$
			and velocity $\beta$,opacity $\kappa$ and the heating index $\alpha$ as labeled. The normalization of the spectra has been performed by using the maximum value of the spectra. }
		\label{fig:normalized_spectra}
	\end{figure*}

	It is shown that for a given velocity, by increasing the ejecta mass $M_{\rm ej}$ and the heating efficiency $f$, the observed temperature increases. From  Fig.~\ref{fig:param_dependence}, it is seen that for higher values of $\beta$, $T_{\rm eff,obs}$  decreases, which is mainly a dynamical effect due to the stronger adiabatic cooling associated with faster expansion, since optically thin regime of the  faster ejecta is achieved at the earlier times ($t_c \propto \beta^{-1}$, Eq.~\ref{critical}). The relativistic corrections, i.e.\ the light-travel-time delay across the equal-arrival-time surface and Doppler boosting, grow with $\beta$ but partially compensate each other.
	For a given velocity, the temperature is nearly insensitive to the heating index at $t_{\rm obs}= 1.43$ days. If the opacity becomes too high, photons are trapped leading to extended adiabatic cooling and resulting in a cooler temperature of the escaped photons. 

		Fig.~\ref{fig:Tobs_comparison} illustrates both the geometrical and dynamical effects on the observed temperature. Here, $T_{\rm obs}(\mu,t_{\rm obs})$ is the observed temperature obtained from Eq. (\ref{e14}), with $T_{\rm emit}$ computed from the dynamical model of Sec.~\ref{sec:thermalevol}.Panel (a) shows the	evolution of $T_{\mathrm{obs}}$ for the line-of-sight direction, $\mu=1$ (solid line) where the shaded region shows the angular spread produced by the equal-arrival-time mapping and Doppler shift introduced in Eq.~(\ref{eq:delay}) and Eq.~(\ref{e14}). For $M_{\rm ej}=0.01\,M_\odot$, $\kappa=0.1~{\rm cm^2\,g^{-1}}$ and $\beta=0.28$ the transparency timescale is $t_{c}=0.95$ days which corresponds to the transition point (described by Eq. (\ref{e15})) shown in panel (a). We also compare the dynamical temperature evolution with a simple power law with decay index $p =(1+\alpha)/4$ (corresponding to the early time behavior of Eq. (\ref{e15})); here we set $\alpha = 0.4$. This result can be compared with the phenomenological prescription presented in some previous studies \citep{Sneppen_2023,Sneppen:2023vkk}, where a simple power law  for  the emitted temperature  $T_{\rm emit}^{\rm PL}(\mu,t_{\rm obs})\propto[t_{\rm obs}(1-\beta\mu)]^{-p}$ is assumed. In those studies the index $p$ was introduced as a constant free parameter; however, this approximation breaks down empirically at later, post-photospheric	epochs, when the temperature declines with different slopes \citep{Sneppen_2024}. It is therefore essential to provide a dynamical description of the thermal evolution that includes the relativistic corrections.
	
	Panel~(b) of Fig.~\ref{fig:Tobs_comparison} shows the ratio 
	$T_{\rm eff,obs}/T_{\rm emit}$ for different ejecta velocities. 
	Here, $T_{\rm eff,obs}$ is the effective temperature inferred from the 
	angularly integrated Doppler-transformed bolometric flux, as defined in 
	Eqs.~(\ref{eee16})--(\ref{eee17}), over the EATS. 
	The reference emitting temperature $T_{\rm emit}$ is obtained from the ejecta evolution 
	by solving the internal energy equations (Eqs.~6, 10, and 11), in the absence of the Doppler and 
	light-travel-time effects, and is evaluated at $t=t_{\rm obs}$, i.e. 
	$T_{\rm emit}=T_{\rm eff}(t_{\rm obs})$. 
	For small $\beta$, the ratio remains close to unity, while at higher velocities, 
	where relativistic corrections become important, $T_{\rm eff,obs}$ deviates significantly 
	from $T_{\rm emit}$; this effect is enhanced at early times.

	Finally, panel~(c) further resolves this angular dependence by showing the density plot of  $T_{\rm obs}(\mu,t_{\rm obs})$ as a function of both observer time and viewing angle. This panel is complementary to panel~(a). As is evident, the observed temperature from  high-latitudes (large $\theta$) has a steeper gradient at early times ($t\leq$ 1 day) while at later times (after the transparency time) we have a more uniform temperature for a broad angular range.

	\begin{figure*}[!htbp]
		\vspace{-2mm} 
		\centering
		\includegraphics[width=.8\linewidth]{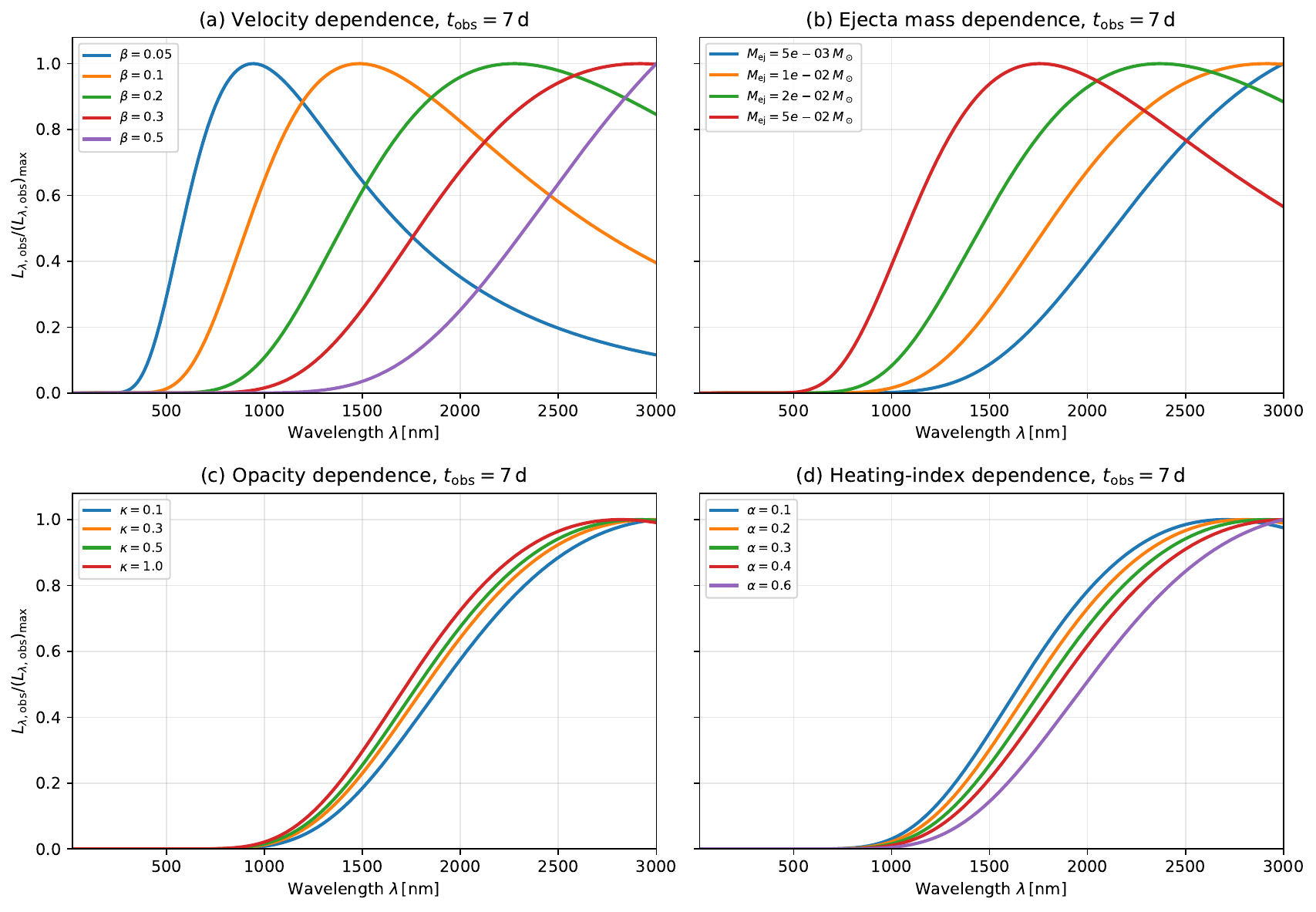}
		\vspace{-1mm} 
		
		\caption{The normalized kilonova spectra for 7 days after merger considering the impact of the  various parameters such as the ejecta mass $M_{ej}$
			and velocity $\beta$, opacity $\kappa$ and the heating index $\alpha$ as labeled. Each spectrum is normalized to its own maximum.}
		\label{fig:normalized_observed_spectra_tobs_7d}
		\vspace{2mm}
	\end{figure*}

	Fig. \ref{fig:Teff_angular_dependence} illustrates how the two  geometrical and relativistic effects introduced above modify the dynamical emitting temperature $T_{\text{emit}}$  (dotted curves) giving rise to  $T_{\mathrm{emit}}(t(\mu,t_{\mathrm{obs}}))$ (dashed curves) obtained from EATS (appears in the numerator of Eq.~\eqref{e14}) and 
	Doppler-corrected observed temperature $T_{\rm obs}(\mu,t_{\rm obs})$ (the solid curves) given by Eq.~\eqref{e14}. Each panel represents the temperature profiles as a function of the viewing angle $\theta$ for various observed times as labeled, different panels showing the result for $\beta_0=0.28$, $0.4$, and $0.8$. Here, we have used  model parameters 
	$\kappa_{\rm avg}=0.022$, 
	$M_{\rm ej}=0.016\,M_\odot$, 
	$\alpha_{\rm avg}=0.166$, and 
	$f=10^{-6}$. By increasing $\theta$,   $T_{\mathrm{emit}}(t(\mu,t_{\mathrm{obs}}))$ is increased, since according to Eq. (\ref{eq:delay}) emission at larger angles originates from earlier times in the rest frame, when the ejecta was hotter, while the line-of-sight emission is released at later times, when the ejecta is cooler. For larger velocities ($\beta_{0}$), the temperature gradient induced by EATS is also enhanced between the emission coming from $\theta =0$ and $\theta=\pi/2$. In contrast, the Doppler boosting gives the opposite temperature trend compared to the light-travel–time delay effect because the Doppler factor increases with $\beta \mu$ (see Eq. (\ref{e14})). At lower velocities these opposing effects largely cancel, so that a single-temperature blackbody spectrum provides an acceptable approximation. As we can see for lower velocities, the emitting temperature  $T_{\text{emit}}$ and the observed temperature $T_{\rm obs}(\mu,t_{\rm obs})$ approximately coincide, while at higher velocities the cancellation of the Doppler and EATS effects does not occur. It is found that the impact of the relativistic corrections on the observed temperature is more significant at early times. It is instructive to compare these results with those obtained in \citep{Sneppen_2023,Sneppen:2023vkk,Sadeh:2025oez}, where it was argued that due to the cancellation  of the relativistic and geometrical effects one can use an almost pure blackbody spectrum to describe the early emission of AT2017gfo. However, it should be noted that for relativistic expansion velocities, and in particular for the early-time emission, the relativistic correction becomes significant, and the deviation of the observed temperature from the emitted temperature at large viewing angles can reach $30\%$.

	Fig. \ref{fig:normalized_spectra} shows the observed spectral luminosity curves, each being normalized with respect to its own maximum at
	$t_{\rm obs}=1.43$ days, obtained by varying one ejecta parameter 
	while keeping the remaining parameters fixed at values
	$\beta=0.3$, $M_{\rm ej}=10^{-2}M_\odot$, $\kappa=0.5~{\rm cm^2\,g^{-1}}$,
	and $\alpha=0.3$. The spectral dependence on the ejecta velocity is presented in panel~(a) for
	$\beta=0.05$--$0.5$. By increasing the velocity, the peak moves towards longer wavelengths and also the spectrum becomes broader, i.e., the luminosity varies more slowly with wavelength.
	
	Panel~(b) shows the effect of varying the ejecta mass in the range $M_{\rm ej}=5\times(10^{-3}-10^{-2})M_\odot$. A larger ejecta mass increases the diffusion timescale and delays the release of the radiation, while from Eqs. (\ref{eq:main}) and (\ref{eq:heating}) we find $L_{peak} \sim \dot{Q}_{\rm in}(t_c)$ which means the peak luminosity increases with the ejecta mass resulting in higher temperatures for the massive ejecta \citep{arnet}. Increasing the temperature for a black body spectrum shifts the spectral peak towards the shorter wavelengths. Panel~(c) presents the opacity dependence for the spectrum. As can be seen, by increasing $\kappa$  
	from $0.1$ to $1.0~{\rm cm^2\,g^{-1}}$ while  the photon transparency time is changed, it slightly modifies the position and width of the spectral peak.

	Finally, panel~(d) shows the dependence on the radioactive heating index $\alpha$. Larger values
	of $\alpha$ correspond to a faster decline of the radioactive energy input and therefore produce a slightly bluer and narrower spectrum.
	Fig. \ref{fig:normalized_spectra} shows that changing the ejecta parameters modifies not
	only the peak position but also the overall shape of the normalized observed spectrum at a fixed observer time. Fig. \ref{fig:normalized_observed_spectra_tobs_7d} extends the comparison shown in Fig.~\ref{fig:normalized_spectra} by illustrating how the relativistic correction evolves with observer time. Seven days after the merger, the peak of each luminosity spectrum shifts toward longer wavelengths. Variations in the ejecta velocity and mass broaden the spectral peaks, whereas changes in $\kappa$ and $\alpha$ shift the peaks outside the plotted wavelength range, showing an effective power-law behavior in the selected wavelength window.
	
	\section{Fitting the Spectra  of AT2017gfo}\label{sec:synthetic}

	\begin{figure*}[!t]
		\centering
		\includegraphics[width=0.8\textwidth]{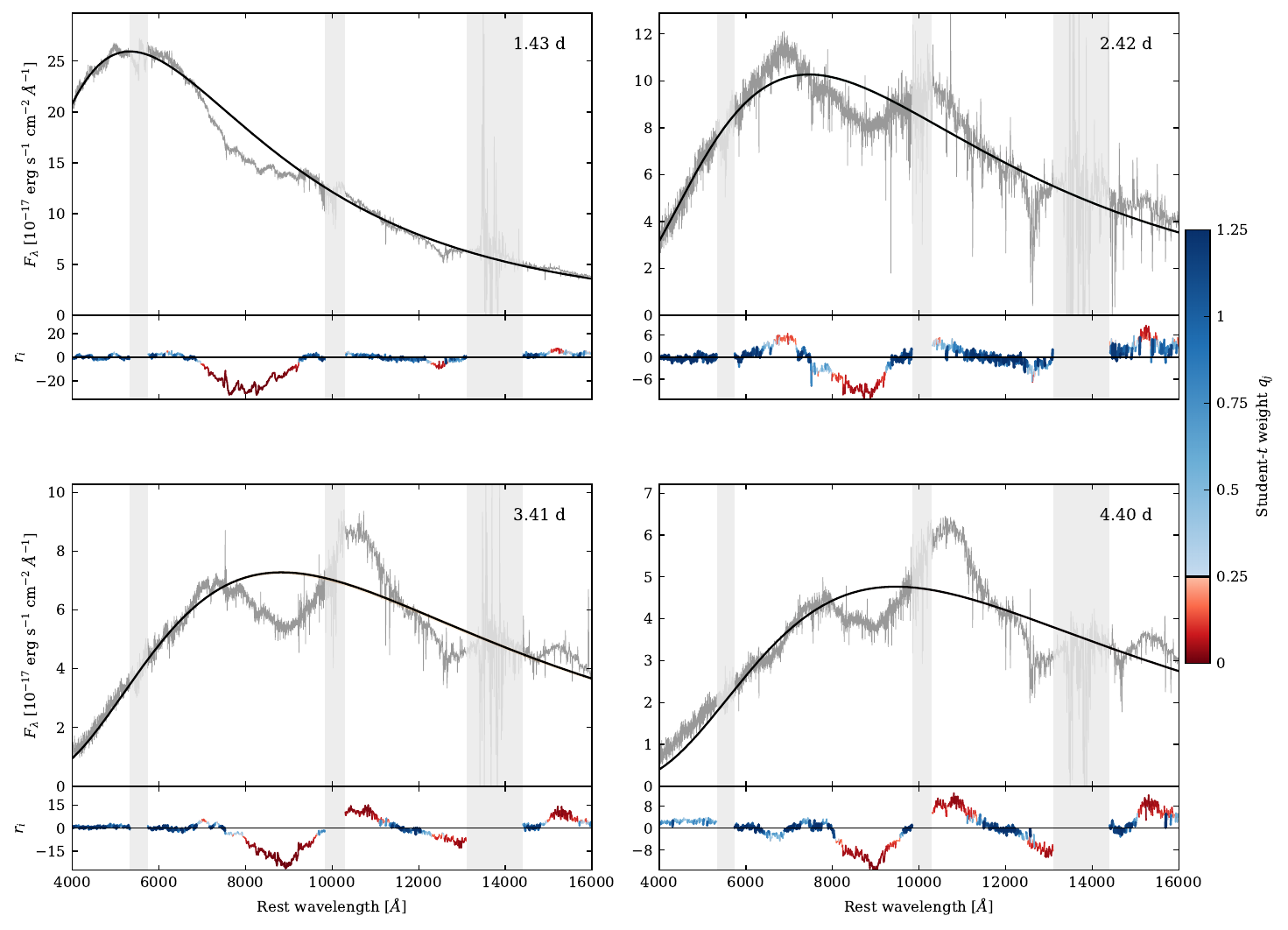}
		\caption{Synthetic spectra from the best-fit dynamical relativistic model compared with the observed spectra of AT2017gfo at four epochs 1.43, 2.42, 3.41, and 4.40 days after the binary merger (upper panels). The lower panels present the corresponding residuals between the model and the data, and the density color shows the down-weighted data points by the $q_i$ parameter. Vertical gray-shaded regions are excluded from the fitting because of systematics within the spectrograph. The parameter values underlying these spectra are summarized in Table \ref{tab:epochfits}.}
		\label{fig:bestfit_spectra}
	\end{figure*}
	The resulting spectra were produced by including the relativistically delayed and Doppler-boosted blackbody emission across the visible photosphere, as described in Sec.~\ref{sec:relativistic}.  The relation between the spectral luminosity $L_\lambda(t_{\rm obs})$ (Eq.(\ref{ee17})) and the flux  $F_{\lambda}$ at a given wavelength,  $L_\lambda(t_{\rm obs})=4 \pi D_{L}^{2} F_{\lambda}$, where $D_L$ is the luminosity distance.  
	In Fig.~\ref{fig:bestfit_spectra}, we fit the X-shooter spectra of AT2017gfo\footnote{\url{http://www.engrave-eso.org/AT2017gfo-Data-Release/}}  at four epochs 1.43, 2.42, 3.41, and 4.40 days after the binary merger, using our dynamical–relativistic framework. The spectra is flux-calibrated, corrected for Galactic extinction, and shifted to the rest frame. Following \citep{Sneppen_2023}, we excluded the wavelength intervals
	5330--5740~\AA, 9840--10300~\AA, and
	13100--14400~\AA\ marked with gray-shaded regions in Fig.~\ref{fig:bestfit_spectra} which are affected by strong
	telluric absorption or noisy overlaps between the UVB, VIS, and NIR arms of the X-shooter spectrograph.

	\begin{deluxetable*}{cccccc}[!t]
		\tablecaption{Posterior constraints for $\beta$, $\kappa M_{\rm ej}$,  $\alpha$ and $D_L$  from the Student-$t$ continuum fits over the four X-shooter epochs. The quoted uncertainties correspond to the 16th--84th percentile interval, here $T_{\rm eff,obs}$ is evaluated from the best fit parameters and $D_L$ is marginalized with the prior of a normal distribution around 40.7 $\pm$ 2.2 Mpc.
			\label{tab:epochfits}}
		\tablehead{
			\colhead{Epoch (days)} &
			\colhead{$\beta$ ($v/c$)} &
			\colhead{$\kappa M_{\rm ej}$ ($10^{-2}\,{\rm cm^2\,g^{-1}\,M_\odot}$)} &
			\colhead{$\alpha$} &
			\colhead{$D_L$ (Mpc)} &
			\colhead{$T_{\rm eff,obs}$ (K)}
		}
		\startdata
		1.43 & $0.199 \pm 0.009$ & $3.8^{+4.0}_{-1.5}$ & $0.49 \pm 0.12$ & $41.5 \pm 2.2$ & $5428 \pm 4$ \\
		2.42 & $0.173 \pm 0.008$ & $4.3^{+4.2}_{-1.9}$ & $0.30 \pm 0.17$ & $41.2 \pm 2.2$ & $3880 \pm 6$ \\
		3.41 & $0.162 \pm 0.007$ & $6.4^{+4.5}_{-2.4}$ & $0.27^{+0.14}_{-0.15}$ & $42.2 \pm 2.1$ & $3278 \pm 6$ \\
		4.40 & $0.122 \pm 0.006$ & $6.5^{+3.9}_{-2.4}$ & $0.22 \pm 0.13$ & $41.6 \pm 2.2$ & $3061 \pm 5$
		\enddata
	
	\end{deluxetable*}
	
	Here, we adopt an independent Student-$t$ likelihood, rather than a Gaussian
	likelihood \citep{Lange01121989}. The primary motivation for adopting a Student-t likelihood is to capture the underlying spectral continuum rather than the detailed absorption and emission features.  The heavier tails of the Student-t likelihood reduce the influence of strong spectral features through the effective weight  $q_{i}=(\nu+1)/(\nu+r_{i}^{2})$  while retaining all spectral points in the analysis. Here, the degrees-of-freedom parameter $\nu$ is set to be 4 and $r_{i}$ is the standardized
	residual  for each retained spectral data point 
	\begin{equation}
		r_i =
		\frac{
			F^{\rm obs}_{\lambda_i}
			-
			F^{\rm mod}_{\lambda_i}
			\left(\hat{\boldsymbol{\theta}}_{\rm MAP}\right)
		}{
			\sigma_i
		},
		\label{eq:qq_standardized_residual}
	\end{equation}
	where $F^{\rm obs}_{\lambda_i}$ is the observed spectral flux,
	$F^{\rm mod}_{\lambda_i}$ is the model flux evaluated at the
	maximum-posterior parameter vector
	$\hat{\boldsymbol{\theta}}_{\rm MAP}=(\kappa,M_{\rm ej},v/c,\alpha,D_{L})$, and $\sigma_i$ is the adopted	statistical uncertainty. The effective weight of the outlier spectral data points decreases as $|r_i|$ increases, and in Fig.~\ref{fig:bestfit_spectra} we highlight in gold the down-weighted data points with $q_i<0.25$, $q_i$ measures  the statistical disagreement with the continuum model.  The down-weighting spectral regions realized by the Student-$t$ likelihood are associated with absorption and emission lines which do not cover by our continuum model. The most prominent region, around $0.70$--$0.97\,\mu$m, overlaps with the Sr\,\textsc{ii} absorption complex \citep{Watson2019,Gillanders2022} and, at later epochs, adjoins the $\sim760$\,nm Y\,\textsc{ii} feature \citep{Sneppen:2023vkk}. Additional low-weight regions coincide with the Sr\,\textsc{ii} emission component near $1.03$--$1.13\,\mu$m \citep{Watson2019,Gillanders2024} and the proposed La\,\textsc{iii}/Ce\,\textsc{iii} absorption around $1.22$--$1.31\,\mu$m \citep{Domoto2022}; the excess beyond $\sim1.5\,\mu$m is at similar wavelengths to the $1.58\,\mu$m NIR emission feature \citep{Gillanders2024}, although it is already present at the first two epochs. 
	The down-weighted points deviate increasingly from the best-fit continuum at later epochs, suggesting that  these structures become stronger relative to the continuum with time.

	For each epoch, the posterior distributions of the ejecta parameters $(\kappa,M_{\rm ej},v/c,\alpha,D_{L})$ were sampled using the affine-invariant ensemble Markov Chain Monte Carlo (MCMC) algorithm implemented in \texttt{emcee} (\citep{Foreman-Mackey2013}). Since $D_L$ sets the overall luminosity normalisation, its uncertainty propagates into the posterior distributions of the inferred continuum properties. Therefore, we marginalize over this uncertainty by adopting a Gaussian prior on $D_L$ \citep{Finstad_2018}, based on the distance measurement of the host galaxy NGC4993, $D_L=40.7\pm2.2~\mathrm{Mpc}$, obtained using $H_0=73.04\pm1.04~\mathrm{km\,s^{-1}\,Mpc^{-1}}$\citep{2022ApJ...934L...7R}. We use 40 walkers for $4\times10^4$--$8.5\times10^4$ steps per walker, discarding the first 30\% of each chain as burn-in.
We adopt uniform priors for $\kappa\in[0.5,3.65]~{\rm cm^2\,g^{-1}}$, $M_{\rm ej}\in[0.023,0.050]~M_\odot$, and  for $\beta\in[0.05,0.5]$ and $\alpha\in[0.02,0.9]$. The radioactive heating follows Eq.~(\ref{eq:heating_rate}), with the normalisation fixed to $f=10^{-6}$.

	We assess convergence using the integrated autocorrelation time $\tau$ \citep{Foreman-Mackey2013}, the split and rank-normalized $\hat R$ statistics \citep{Vehtari2021}, and by comparing the posterior medians of the two halves of the retained chains. The retained chains span $\geq24\,\tau_{\max}$ post-burn-in steps, where $\tau_{\max}$ is the largest autocorrelation time among the sampled parameters, yielding effective sample sizes $\gtrsim9\times10^2$ for all parameters.  The rank-normalized $\hat R$ is $\leq1.03$ for all parameters at $t\ge2.42$\,d and $\leq1.06$ at 1.43\,d, remaining below the commonly adopted threshold of $\hat R<1.1$, with the 1.43\,d chains showing the weakest convergence. The posterior medians of the two chain halves agree to within $0.1\sigma$ in all cases.

	\begin{figure}[!t]
		\centering
		\includegraphics[width=\columnwidth]{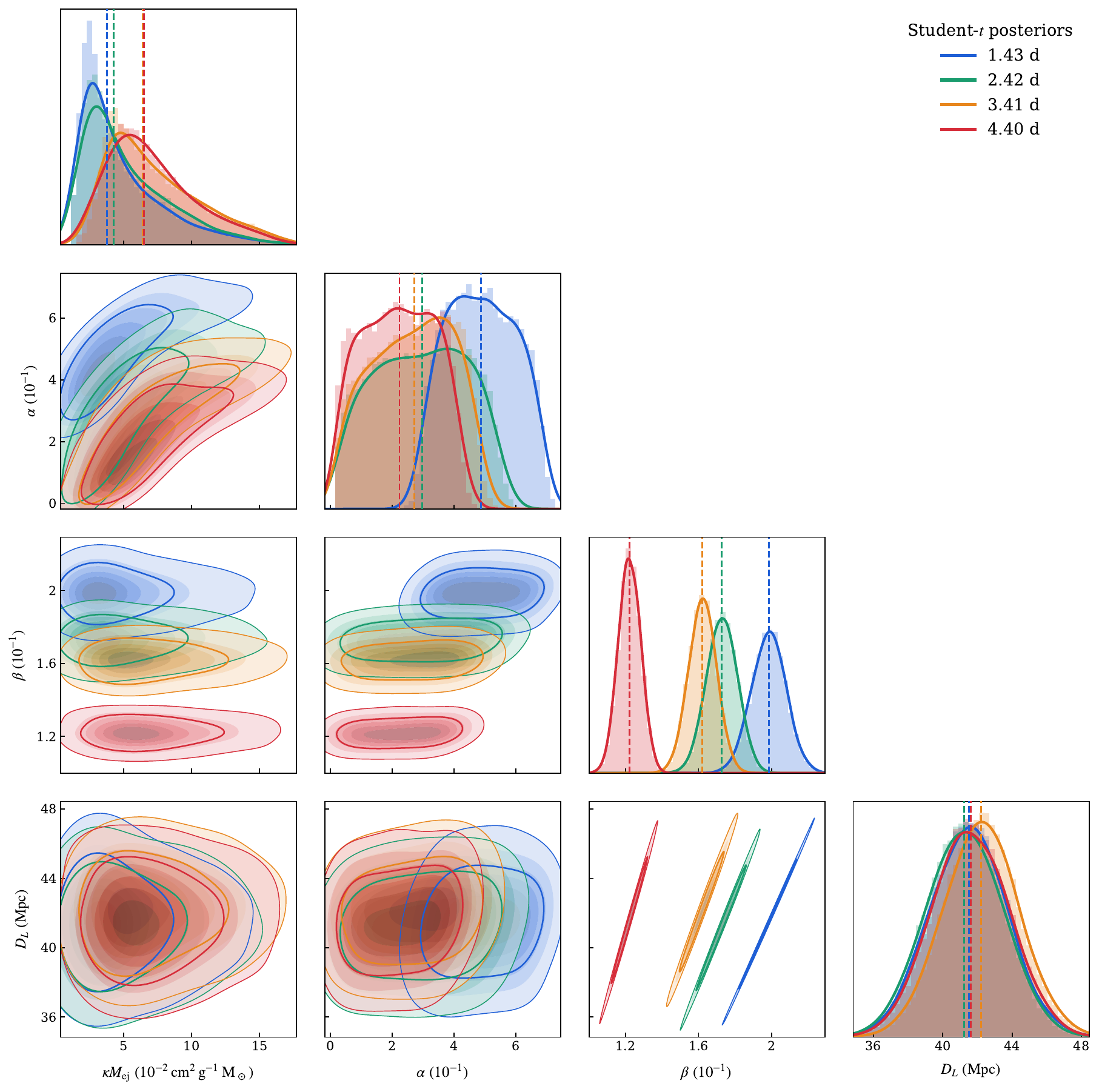}
		\caption{Posterior distributions of the Student-$t$ continuum fits for AT2017gfo at four epochs after merger (1.43, 2.42, 3.41, and 4.40 days).}
		\label{fig:posterior_overlay}
	\end{figure}
	
	The posterior distributions for all four epochs are shown together in Fig.~\ref{fig:posterior_overlay}, and the inferred parameters are listed in Table~\ref{tab:epochfits}. Due to the intrinsic degeneracy between $\kappa$ and $M_{\rm ej}$, previously noted by \citet{Li_2019}, we present the results in terms of the effective combination $\kappa M_{\rm ej}$. We see that, the velocity parameter $\beta$ decreases as the time passes, from $\simeq 0.20 $ at 1.43~days to $\simeq0.12$ at 4.40~days, with uncertainties marginalized over the luminosity distance. Since each epoch is fitted
	independently with our single-zone model, $\beta$ is an effective parameter, and its decrease between the epochs may be interpreted as the effective emitting surface receding into slower layers of the homogeneously expanding ejecta. Although the $\kappa M_{\rm ej}$ posteriors of the four epochs overlap substantially, their medians increases mildly, from
	$(3.8^{+4.0}_{-1.5})\times10^{-2}$ to
	$(6.5^{+3.9}_{-2.4})\times10^{-2}\,{\rm cm^{2}\,g^{-1}\,M_\odot}$ between 1.43 and 4.40~days, it may reflect a rising effective opacity as the ejecta cools, consistent with the spectral reddening in Fig.~\ref{fig:bestfit_spectra}.
	Breaking the degeneracy between $\kappa$ and $M_{\rm ej}$ requires a multi-epoch treatment or complementary data.
	The observed effective temperature $T_{\rm eff,obs}$ in
	Table~\ref{tab:epochfits} is computed from the best-fit parameters of each epoch through Eqs.~(\ref{eee16})--(\ref{eee17}). It decreases from $5428\pm4$~K at
	1.43~d to $3061\pm5$~K at 4.40~d, consistent with the progressive reddening of the spectra in Fig.~\ref{fig:bestfit_spectra}.
	
To quantify the impact of neglecting relativistic transfer effects, we repeated the fits of all four epochs within the same dynamical framework, but without Doppler boosting and light-travel-time effects. Neglecting these effects increases the inferred dynamical temperature by up to $7.6\%$ and the velocity by $\simeq 11.9\%$ for these epochs, in agreement with the results reported by \citet{Sadeh:2025oez}.
 The residuals in Fig.~\ref{fig:bestfit_spectra} are concentrated in specific wavelength ranges and most likely arise from wavelength-dependent opacities, which are not captured by our grey treatment. Individual atomic species can produce localized absorption and enhanced opacity at particular wavelengths \citep{Kasen:2013,Tanaka:2013,Watson2019}, and departures from local thermodynamic equilibrium (non-LTE effects) can further modify the spectral shape \citep{Pognan:2022}. We will address these effects in future work.

    According to Figs. \ref{fig:normalized_spectra} and \ref{fig:normalized_observed_spectra_tobs_7d}, in the case of more rapid ejecta, we would expect to see the reddening and broadening of the spectrum by the significant contribution of the relativistic and geometrical effects. The recent observation of AT2025ulz \citep{Hall:2025qsm},  shows unusual rapid fading and substantial reddening over the first 5 days before plateauing and subsequently rising and it shows some deviations from AT2017gfo which may indicate the necessity of including relativistic effects in this event.

	\section{Conclusion}
	\label{sec:conclusion}
	
	In this work, we developed a dynamical--relativistic framework for the modeling of kilonova emission and applied it to the spectra of the AT2017gfo event. In our model, the thermal evolution of the expanding ejecta is described dynamically by solving the internal-energy equation under the assumptions of homologous expansion, radioactive heating, adiabatic cooling, and radiative diffusion. The dynamical model of \citep{Li_2019} included the physical thermal evolution but did not incorporate the relativistic and geometrical effects required for rapidly expanding kilonova ejecta. On the other hand, the relativistic framework presented in \citep{Sneppen_2023,Sneppen:2023vkk} showed that the combined EATS and Doppler corrections can modify the observed spectra, but the emitted temperature evolution was prescribed phenomenologically as a power law. The present work bridges these two approaches by using a self-consistent dynamical temperature evolution as a function of the ejecta properties, such as the ejecta mass, opacity, expansion velocity, heating efficiency, and radioactive heating index and then applying relativistic corrections to generate the observed spectrum.

A key outcome of this approach is that the emitted effective temperature follows a broken power-law evolution, which emerges naturally from the transition between the optically thick and optically thin ejecta at the transparency timescale. We examined the effect of the ejecta parameters on the thermal evolution of the emitted and observed radiation and, subsequently, on the spectral luminosity. We found that the ejecta velocity affects both the relativistic corrections and the dynamical evolution of the ejecta. The observed temperature is shaped by a competition between the geometric time delay and Doppler boosting, and at lower velocities these effects largely cancel each other. As a result, the emission at lower velocities is well described by a single-temperature blackbody, in agreement with \citet{Sneppen_2023}, whereas for faster expanding ejecta, especially at $\beta \gtrsim 0.5$, stronger angular temperature gradients are produced, leading to more significant spectral distortions. Even for moderate velocities, $\beta \simeq 0.12$--$0.30$, neglecting Doppler boosting and light-travel-time effects leads to an overestimation of the dynamical temperature and the velocity by up to $7.6\%$ and $\simeq 11.9\%$, respectively.

We also fitted the early X-shooter spectra of AT2017gfo at 1.43, 2.42, 3.41, and 4.40~days within this framework, using a Student-$t$ likelihood to capture the underlying continuum while limiting the impact of individual absorption and emission features, and marginalizing over the distance to NGC~4993 (Table~\ref{tab:epochfits}). Between the first and last epochs, the effective velocity decreases from $\beta\simeq0.20$ to $\simeq0.12$ and the observed effective temperature from $\simeq5400$ to $\simeq3100$~K, while the combination $\kappa M_{\rm ej}$ increases mildly from $\simeq3.8\times10^{-2}$ to $\simeq6.5\times10^{-2}\,{\rm cm^2\,g^{-1}}\,M_\odot$. The decrease in the inferred velocity of the ejecta and the increase in $\kappa M_{\rm ej}$ (i.e., a higher effective opacity) toward later epochs may hint at the appearance of a red component. Such a component has been reported previously in multi-band, bolometric light-curve and spectral analyses of AT2017gfo \citep{Cowperthwaite2017,2017Sci...358.1570D,Kasen_2017,Perego_2017,Villar2017}, in which the early emission is dominated by fast, lanthanide-poor ejecta and the later emission by slower, lanthanide-rich material. However, continuum fits alone cannot unambiguously determine whether this trend arises from a change in opacity, in ejecta mass, or in the dominant emitting component, and because of the $\kappa$--$M_{\rm ej}$ degeneracy only their product is constrained.

	Here, we adopted the diffusion-limit approximation, which is most reliable
	at early times, while deviations are expected at later times when the ejecta
	become more transparent and the radiation field departs from LTE. We also
	used a grey, wavelength-independent opacity, which prevents the model from
	capturing detailed line features, localized absorption structures, and the
	wavelength-dependent influence of r-process elements \citep{Kasen:2013,Tanaka:2013,2019Natur.574..497W}. The residuals in the
	spectral fits are therefore likely connected to missing wavelength-dependent
	opacity, line transfer, and possible non-LTE effects \citep{Pognan:2022}, which will be the
	subject of our future work. Such an extension would connect the continuum evolution more
	directly to the ejecta composition and to the contribution of individual atomic species. The inferred uncertainties are further subject to systematics from the single-zone treatment and from observational effects such as flux calibration and telluric correction, which can correlate neighboring wavelengths. 
	
	A more complete treatment should also go beyond the simple diffusion-limit description and include detailed radiative transfer through a multi-frequency opacity structure. This would provide a fully time-dependent and wavelength-dependent dynamical--radiative model capable of interpreting both the light curve and spectral evolution of future kilonovae, especially events with higher expansion velocities, where relativistic effects are expected to be more significant. 
   Such fast ejecta are physically expected: numerical simulations of binary neutron star mergers predict a fast tail of dynamical ejecta with velocities $\gtrsim 0.6c$ \citep{Hotokezaka:2013,Metzger:2015,Radice:2018}, and the late-time X-ray excess of GW170817 has been interpreted as the afterglow of such fast-moving material \citep{Hajela:2022}.  Kilonovae with more energetic outflows, observed at earlier times or from different viewing angles than AT2017gfo, including those accompanying long-duration gamma-ray bursts \citep{Rastinejad:2022,Levan:2024}, are therefore natural cases in which relativistic and equal-arrival-time effects must be included to infer the ejecta properties reliably. Despite its simplifying assumptions, the presented model captures key features of the relativistic dynamical evolution of kilonova emission, which may shed light on the nature of these transients.

	\section*{{Acknowledgments}}
	The authors are deeply grateful to Siamak Sadat Gousheh and Albert Sneppen
	for the careful reading   of the manuscript and the insightful comments  which improved the quality of the paper. We also appreciate the valuable discussions with Parisa Hashemi at the initial phase of the project and also her assistance with the numerical fitting.

\end{document}